\documentclass[aps,prb,showpacs,twocolumn,citeautoscript,superscriptaddress]{revtex4}
\usepackage{amsmath}
\usepackage{graphicx}
\usepackage{dcolumn}
\usepackage{float}

\begin{document}
\title{Ferromagnetic Ordering in CeIr$_{2}$B$_{2}$: Transport, magnetization, specific heat and NMR studies}
\author{A. Prasad}
\affiliation{Department of Physics, Indian Institute of Technology, Kanpur 208016, India}
\author{V. K. Anand}
\email{vivekkranand@gmail.com}
\affiliation{Department of Physics, Indian Institute of Technology, Kanpur 208016, India}
\affiliation{Ames Laboratory and Department of Physics and Astronomy, Iowa State University, Ames, Iowa 50011, USA}
\author{U. B. Paramanik}
\author{Z. Hossain}
\email{zakir@iitk.ac.in}
\affiliation{Department of Physics, Indian Institute of Technology, Kanpur 208016, India}
\author{R. Sarkar}
\author{N. Oeschler}
\author{M. Baenitz}
\author{C. Geibel}
\affiliation {Max-Planck Institute for Chemical Physics of Solids, 01187 Dresden, Germany}

\date{\today}

\begin{abstract}

We present a complete characterization of ferromagnetic system CeIr$_{2}$B$_{2}$ using powder x-ray diffraction XRD, magnetic susceptibility $\chi(T)$, isothermal magnetization $M(H)$, specific heat $C(T)$, electrical resistivity $\rho(T,H)$, and thermoelectric power $S(T)$ measurements. Furthermore $^{11}$B NMR study was performed to probe the magnetism on a microscopic scale. Rietveld refinement of powder XRD data confirms that CeIr$_{2}$B$_{2}$ crystrallizes in CaRh$_{2}$B$_{2}$-type orthorhombic structure (space group \textit{fddd}). The $\chi(T)$, $C(T)$ and $\rho$(T) data confirm bulk ferromagnetic ordering with $T_c = 5.1$~K. Ce ions in CeIr$_{2}$B$_{2}$ are in stable trivalent state. Our low-temperature $C(T)$ data measured down to 0.4~K yield Sommerfeld coefficient $\gamma$ = 73(4)~mJ/mol\,K$^{2}$ which is much smaller than the previously reported value of $\gamma$ = 180~mJ/mol\,K$^{2}$ deduced from the specific heat measurement down to 2.5~K\@. For LaIr$_{2}$B$_{2}$ $\gamma$ = 6(1)~mJ/mol\,K$^{2}$ which implies the density of states at the Fermi level ${\cal D}(E_{F}) = 2.54$~states/(eV f.u.) for both spin directions. The renormalization factor for quasi-particle density of states and hence for quasi-particle mass due to $4f$ correlations in CeIr$_{2}$B$_{2}$ is $\approx 12$. The Kondo temperature $T_{\rm K}\sim 4$~K is estimated from the jump in specific heat of CeIr$_{2}$B$_{2}$ at $T_c$. Both $C(T)$ and $\rho(T)$ data exhibit gapped-magnon behavior in magnetically ordered state with an energy gap $E_g \sim 3.5$~K\@. The $\rho$ data as a function of magnetic field $H$ indicate a large negative magnetoresistance (MR) which is highest for $T=5$~K. While at 5~K the negative MR keeps on increasing up to 10~T, at 2~K an upturn is observed near $H = 3.5$~T\@. On the other hand, the thermoelectric power data have small absolute values ($S \sim 7~\mu$V/K) indicating a weak Kondo interaction. A shoulder in $S(T)$ at about 30 K followed by a minimum at $\sim$~10~K is attributed to crystal electric field (CEF) effects and the onset of magnetic ordering. $^{11}$B NMR line broadening provides strong evidence of ferromagnetic correlations below 40~K.

\end{abstract}


\maketitle

\section{\label{Intro} INTRODUCTION}

Rare earth based intermetallic compounds have drawn a great interest in last few decades due to their divergent unusual physical properties in a variety of crystal structures. Among these, Ce-based compounds are well known for exhibiting a wide range of physical properties such as long range magnetic ordering with different ground states, Kondo lattice behavior, heavy fermion behavior, valance fluctuation, superconductivity and quantum criticality which arise from the competition between the RKKY and the Kondo interactions.\cite{Varma, Georges, Riseborough, Stewart, Amato, Lohneysen} The family of CeT$_{2}$X$_{2}$ (T= transition element, X = Ge, Si) compounds have been intensively studied because of their interesting magnetic and superconducting properties. \cite{Steglich, Jaccard, Felten, Gupta, Knopp, Berisso} For example, Kondo lattice antiferromagnet CeCu$_{2}$Si$_{2}$ is a heavy fermion superconductor that exhibits the coexistence of magnetism and superconductivity.\cite{Steglich, Steglich1996} Pressure studies on partially Ge-substituted CeCu$_{2}$Si$_{2}$ reveal two distinct superconducting phases associated with two quantum critical points.\cite{Yuan} Heavy fermion antiferromagnet CePd$_{2}$Si$_{2}$ is another interesting system which exhibit quantum critical behaviour that can be tuned by pressure and presents an intriguing phase diagram: a pressure induced superconductivity is observed around the critical pressure ($\sim 28$~kbar) together with a non-Fermi liquid behaviour after the complete suppression of antiferromagnetism.\cite{Grier, vanDijk, Grosche, Demuer}

The CeT$_{2}$X$_{2}$ compounds are very sensitive to changes in unit cell volume. Even a small change in volume can lead to a drastic change in structural as well as physical properties, for example, Kondo lattice system CeIr$_{2}$Ge$_{2}$ which forms in CaBe$_{2}$Ge$_{2}$-type tetragonal structure exhibits heavy fermion behavior, \cite{Mathur,Sampatkumaran1996, Mallik} and valence fluctuating system CeIr$_{2}$Si$_{2}$ which shows polymorphism exhibits Fermi liquid behavior in low temperature ThCr$_{2}$Si$_{2}$-type tetragonal structure and non-Fermi liquid behavior in high temperature CaBe$_{2}$Ge$_{2}$-type tetragonal structure. \cite{Hiebl, Mihalik} Most of the CeT$_{2}$X$_{2}$ compounds have been found to crystallize in layered ThCr$_{2}$Si$_{2}$-type body-centered tetragonal structure which is very favorable for magnetic ordering and superconductivity on account of its centrosymmetric nature. However, there are few compounds that form in different crystal structure. CeIr$_{2}$Ge$_{2}$ and CeIr$_{2}$Si$_{2}$ are such examples.  Another such compound is CeIr$_{2}$B$_{2}$.

\begin{figure}
\includegraphics[width=8cm, keepaspectratio]{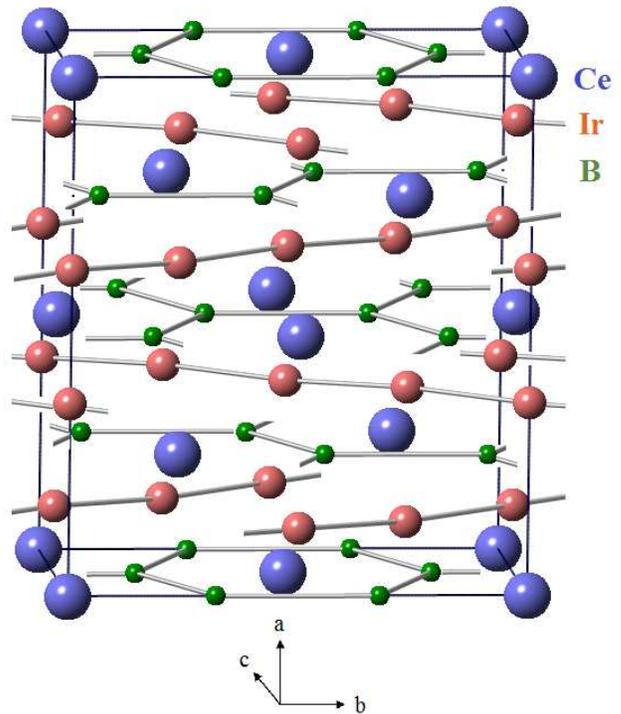}
\caption{\label{fig:CeIr2B2_structure} (Color online) The CaRh$_{2}$B$_{2}$-type orthorhombic structure (space group \textit{fddd}) of CeIr$_{2}$B$_{2}$. For clarity, the origin of the unit cell shown is shifted by (1/8, 1/8, 1/8) from the atomic cordinates listed in Table~\ref{tab:XRD}.}
\end{figure}

CeIr$_{2}$B$_{2 }$ is reported to crystallize in an orthorhombic structure (space group \textit{fddd})\cite{Jung} in contrast to the tetragonal ThCr$_{2}$Si$_{2}$ structure for most compounds with 1:2:2 ratio. The CaRh$_{2}$B$_{2}$-type orthorhombic structure of CeIr$_{2}$B$_{2 }$ is shown in Fig.~\ref{fig:CeIr2B2_structure}. As can be seen from Fig.~\ref{fig:CeIr2B2_structure}, this orthorhombic structure is quite different from the tetragonal ThCr$_{2}$Si$_{2}$ structure. In this structure Ce atoms lie at the center of a quasi-hexagonal network (in crystallographic $b$-$c$ plane) formed by B atoms, while Ir atoms form slightly bent crossed chains as shown in Fig.~\ref{fig:CeIr2B2_structure}. The layers of Ce-B hexagons and Ir chains are stacked along the $a$-axis.

Based on investigations by magnetic susceptibility (2--300~K), specific heat (2.5--20~K) and electrical resistivity (1.4--300~K) measurements, CeIr$_{2}$B$_{2}$ is reported to undergo a ferromagnetic ordering below 6 K.\cite{Sampatkumaran} The Sommerfeld coefficient ($\gamma$) was estimated to be 180~mJ/mol\,K$^{2}$ and this large value was attributed to the possible heavy fermion behavior in this compound.\cite{Sampatkumaran}

Recently we investigated the physical properties of a related compound PrIr$_{2}$B$_{2}$ which also forms in the CaRh$_{2}$B$_{2}$-type orthorhombic structure. \cite{Anupam} We found evidences of spin-glass behavior in PrIr$_{2}$B$_{2}$ which is very exciting considering that there are only few stoichiometric and crystallographically well ordered intermetallic compounds known to exhibit spin-glass behavior. Though the mechanism behind the spin-glass behavior in PrIr$_{2}$B$_{2}$ is not clear at the moment, the presence of a CEF-split singlet ground state as inferred from the specific heat data indicates that possibly the underlying mechanism lies in the dynamic fluctuations of the low-lying crystal field levels.

Thus, we see that the RIr$_{2}$B$_{2}$ (R = rare earths) compounds present interesting avenue to explore the novel electronic ground state properties arising from the competing RKKY, Kondo and CEF interactions. Considering the fact that there are not many known ferromagnetically ordered heavy fermion systems, we decided to investigate the low temperature properties of CeIr$_{2}$B$_{2}$.  Our detailed investigations of low temperature properties of CeIr$_{2}$B$_{2}$ by magnetic susceptibility $\chi(T)$, isothermal magnetization $M(H)$, specific heat $C(T)$, electrical resistivity $\rho(T)$, magnetoresistance $\Delta\rho(H)/\rho(0)$, thermoelectric power $S(T)$ and $^{11}$B NMR measurements down to 0.4~K confirm the long range ferromagnetic ordering, $T_c =  5.1$~K, however the value of $\gamma$ ($= 73(4)$~mJ/mol\,K$^{2}$) is found to be not as enhanced as reported previously. Further, we also observe a gapped-magnon behavior in magnetically ordered state in both $C(T)$ and $\rho(T)$ data which indicate the magnetic properties of this compound to be highly anisotropic. The thermoelectric power data indicate a weak Kondo interaction, and the line broadening of $^{11}$B NMR spectra provides evidence of ferromagnetic correlations below 40~K which is well above the ordering temperature.

\section{\label{ExpDetails} EXPERIMENTAL DETAILS}

We prepared polycrystalline sample of CeIr$_{2}$B$_{2}$ using the standard arc-melting technique. High purity elements (99.9\% and above) of Ce, Ir, and B were taken in the stoichiometric 1:2:2 ratio and melted for 4-5 times in the vacuum chamber under ultra high purity argon atmosphere. Additional amount of B was added to compensate the weight loss by flying-off of B. In order to homogenize and improve the reaction among the constituents, we annealed the arc-melted sample at 1200 $^\circ$C for a week. The phase purity of the annealed sample was checked by powder x-ray diffraction (XRD) using Cu $K_\alpha$ radiation and scanning electron microscope (SEM). The chemical composition of the sample was checked by energy dispersive x-ray (EDX) analyzer attached with SEM.

Magnetization measurements were carried out using a commercial superconducting quantum interference device (SQUID) magnetometer (MPMS, Quantum-Design). Electrical resistivity, specific heat and thermoelectric power measurements were carried out using the AC transport, heat capacity and thermal transport options, respectively, of a physical properties measurement system (PPMS, Quantum-Design).  $^{11}$B field sweep NMR measurements were performed with a standard pulsed NMR spectrometer (Tecmag) at the fixed frequency of 105 MHz and in the temperature range of 4.2--295~K. The Knight shift $^{11}K$ was measured with respect to the reference compound H$_{3}$BO$_{3}$ with $^{11}K = 0$.

\begin{figure}
\includegraphics[width=8.5cm, keepaspectratio]{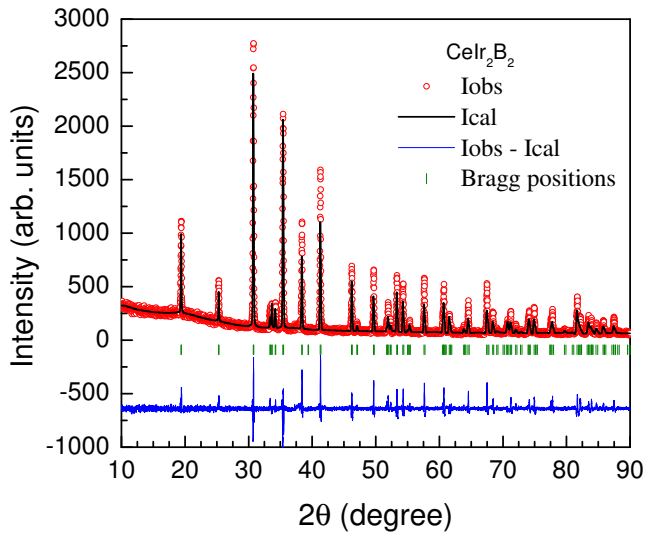}
\caption{\label{fig:CeIr2B2_XRD} (Color online) Powder x-ray diffraction patterns of CeIr$_{2}$B$_{2}$ recorded at room temperature. The solid line through the experimental points is the Rietveld refinement profile calculated for the CaRh$_{2}$B$_{2}$-type orthorhombic structure (space group \textit{fddd}). The short vertical bars mark the Bragg peak positions. The lowermost curve represents the difference between the experimental and calculated intensities.}
\end{figure}

\begin{table}
\caption{\label{tab:XRD} Crystallographic parameters obtained from the structural Rietveld refinement of powder XRD data of CeIr$_2$B$_2$. The refinement quality parameter $\chi ^2= 3.57$.}
\begin{ruledtabular}
\begin{tabular}{llll}

Structure &\multicolumn{3}{l} {CaRh$_{2}$B$_{2}$-type orthorhombic} \\
Space group & \textit{Fddd} \\
Formula units/unit cell (Z)  &  8 \\
\multicolumn{2}{l}{Lattice  parameters} \\
 \hspace{1cm} $a$ (\AA)  & 10.6476(4)  \\
 \hspace{1cm} $b$ (\AA)  & 9.3805(4)   \\
 \hspace{1cm} $c$ (\AA)  & 6.0222(3)  \\
 \hspace{1cm} $V_{cell}$ (\AA$^3$)& 601.50(5) \\
 &\\
Refined Atomic Coordinates\\
\hspace{0.7cm}Atom \hspace{0.5cm} Wyckoff & x &y & z  \\
\hspace{1cm}Pr \hspace{1cm} 8a & 1/8 & 1/8 & 1/8 \\
\hspace{1cm}Ir \hspace{1cm} 16e & 0.4957(2) & 1/8 & 1/8 \\
\hspace{1cm}B  \hspace{1cm} 16f & 1/8 & 0.461(6) & 1/8\\

\end{tabular}
\end{ruledtabular}
\end{table}

\section{\label{Results} RESULTS AND DISCUSSION}

Powder x-ray diffraction data were collected on the crushed sample and analyzed by Rietveld refinement using {\tt FullProf} software.\cite{Rodriguez} Rietveld refinement confirmed the CaRh$_{2}$B$_{2}$-type orthorhombic structure (space group \textit{fddd}) of CeIr$_{2}$B$_{2}$. The refinement reveal the single phase nature of the sample without any impurity peak. The room temperature XRD pattern and Rietveld refinement profile are shown in Fig.~\ref{fig:CeIr2B2_XRD}. The crystallographic and refinement quality parameters are listed in Table~\ref{tab:XRD}. The lattice parameters $a = 10.6476(4)$~{\AA}, $b = 9.3805(4)$~\AA\ and $c = 6.0222(3)$~\AA\ of CeIr$_{2}$B$_{2}$ are in good agreement with the reported values $a = 10.645$~{\AA}, $b = 9.379$~{\AA}, $c = 6.019$~\AA\@.\cite{Jung} The EDX composition analysis indicated the Ce:Ir ratio to be close to 1:2, however the B-content could not be determined precisely from EDX analysis. The high resolution SEM images also confirmed the single phase nature of the sample; no noticeable impurity phase was observed in SEM images.

\begin{figure}
\includegraphics[width=8cm, keepaspectratio]{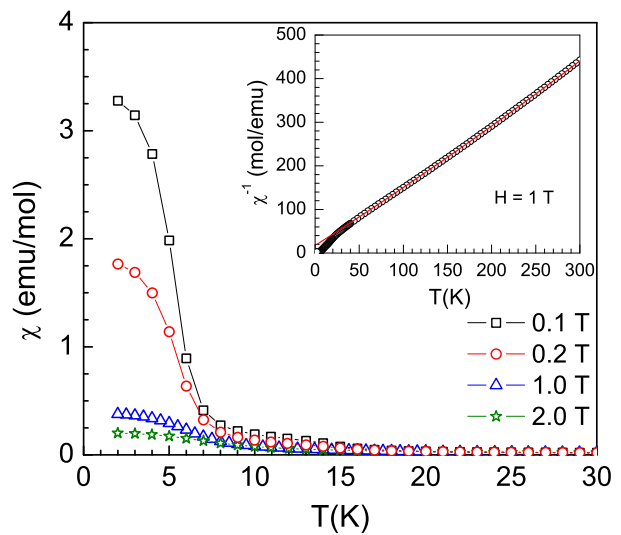}
\caption{\label{fig:CeIr2B2_M-T} (Color online) Temperature dependence of magnetic susceptibility data, $\chi (T)$ of CeIr$_{2}$B$_{2}$ at different applied fields, $H = 0.1,~0.2,~1~{\rm and}~2$~T\@. The inset shows the inverse susceptibility $\chi^{-1}(T)$ plot for $H$ = 1 T. The solid line in inset represents the fit to modified Curie-Weiss behavior.}
\end{figure}

The temperature $T$ dependence of the magnetic susceptibility $\chi$ of CeIr$_{2}$B$_{2}$ is shown in Fig.~\ref{fig:CeIr2B2_M-T} for different applied magnetic fields. At low applied fields, e.g., at $H = 0.1$~T, a steep increase is observed at 6 K in $\chi (T)$ which tends to saturate below 4~K, evidencing the long range ferromagnetic ordering. An increase in applied field results in a decrease in magnitude of $\chi$ at low temperature as expected for ferromagnetic ordering. At high temperature the $\chi (T)$ data follows the modified Curie-Weiss behavior, $\chi = \chi_0 + C/(T - \theta_{\rm p})$. The inverse susceptibility plot $\chi^{-1}(T)$ for $\chi (T)$ data measured at $H = 1$~T is shown in the inset of Fig.~\ref{fig:CeIr2B2_M-T}. A fit of $\chi^{-1}(T)$ data by modified Curie-Weiss behavior in the temperature range 50--300~K (shown by the solid curve in the inset of Fig.~\ref{fig:CeIr2B2_M-T}) yields temperature independent susceptibility $\chi_0 = -2.15(8) \times 10^{-4}$~emu/mol, Curie constant $C = 0.779(3)$~emu\,K/mol and the Weiss temperature $\theta_{\rm p} = -13.4(2)$~K\@. The effective magnetic moment $\mu_{\rm eff}$ calculated from the value of $C$ is, $\mu_{\rm eff} = 2.50(1)\,\mu_{B}$ which is very close to the theoretically expected value of $2.54\,\mu_{B}$ for Ce$^{3+}$ ions. Interestingly we observe that both $\chi_0$ and $\theta_{\rm p}$ are negative. A negative $\theta_{\rm p}$ suggests the presence of antiferromagnetic correlations. A neutron diffraction study is called for to understand the magnetic structure of this compound.

\begin{figure}
\includegraphics[width=8cm, keepaspectratio]{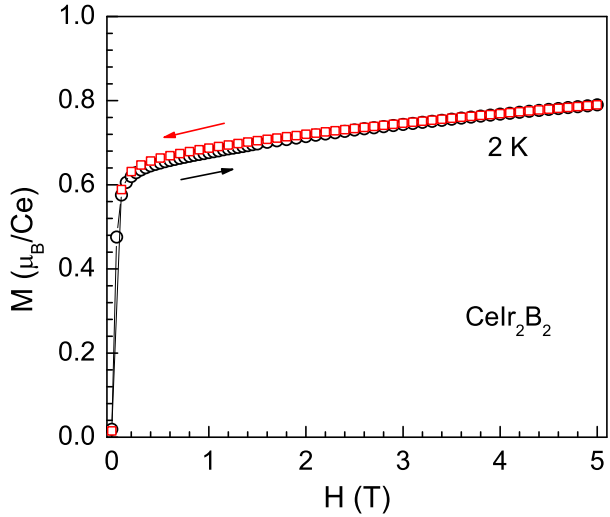}
\caption{\label{fig:CeIr2B2_M-H} (Color online) Magnetic field dependence of isothermal magnetization data, $M(H)$ of CeIr$_{2}$B$_{2}$ measured at constant temperature 2 K.}
\end{figure}

In Fig.~\ref{fig:CeIr2B2_M-H}, the isothermal magnetization $M$ is displayed as a function of the applied field $H$ at 2 K.  As expected for a ferromagnetic system, the $M(H)$ data show a rapid increase at very low field, reaching a value of 0.65~$\mu_{B}$/Ce for $H = 0.2$~T\@. This is followed by a weak linear increase of $M$ reaching a value $\sim 0.8~\mu_{B}$/Ce at 2~K and 5~T. The hysteresis in $M(H)$ curves measured during increasing and decreasing cycles of magnetic field is almost negligible.

\begin{figure}
\includegraphics[width=8cm, keepaspectratio]{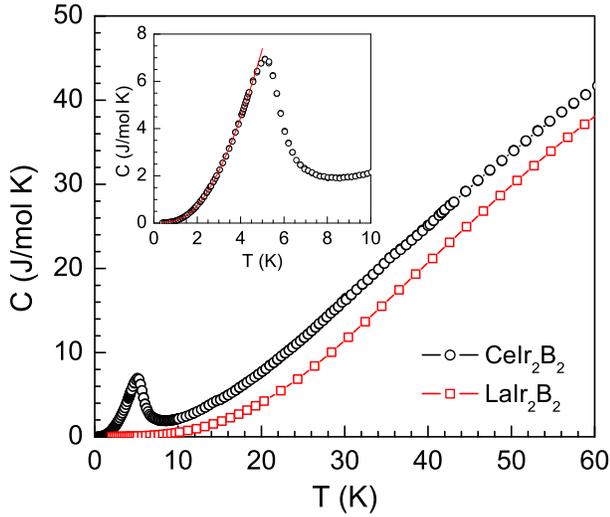}
\caption{\label{fig:CeIr2B2_HC} (Color online) Temperature dependence of the specific heat data, $C(T)$ of CeIr$_{2}$B$_{2}$ measured in zero field in the temperature range (0.4--60~K). The $C(T)$ data of nonmagnetic reference compound LaIr$_{2}$B$_{2}$ (Ref.~\onlinecite{Anupam}) are also shown. Inset shows an expanded view of low temperature $C(T)$ data. The solid curve in inset represents a fit for gapped-magnon behavior by Eq.~(\ref{eq:C-magnon}) for $0.4~{\rm K} \leq T \leq 4.5$~K.}
\end{figure}

\begin{figure}
\includegraphics[width=8cm, keepaspectratio]{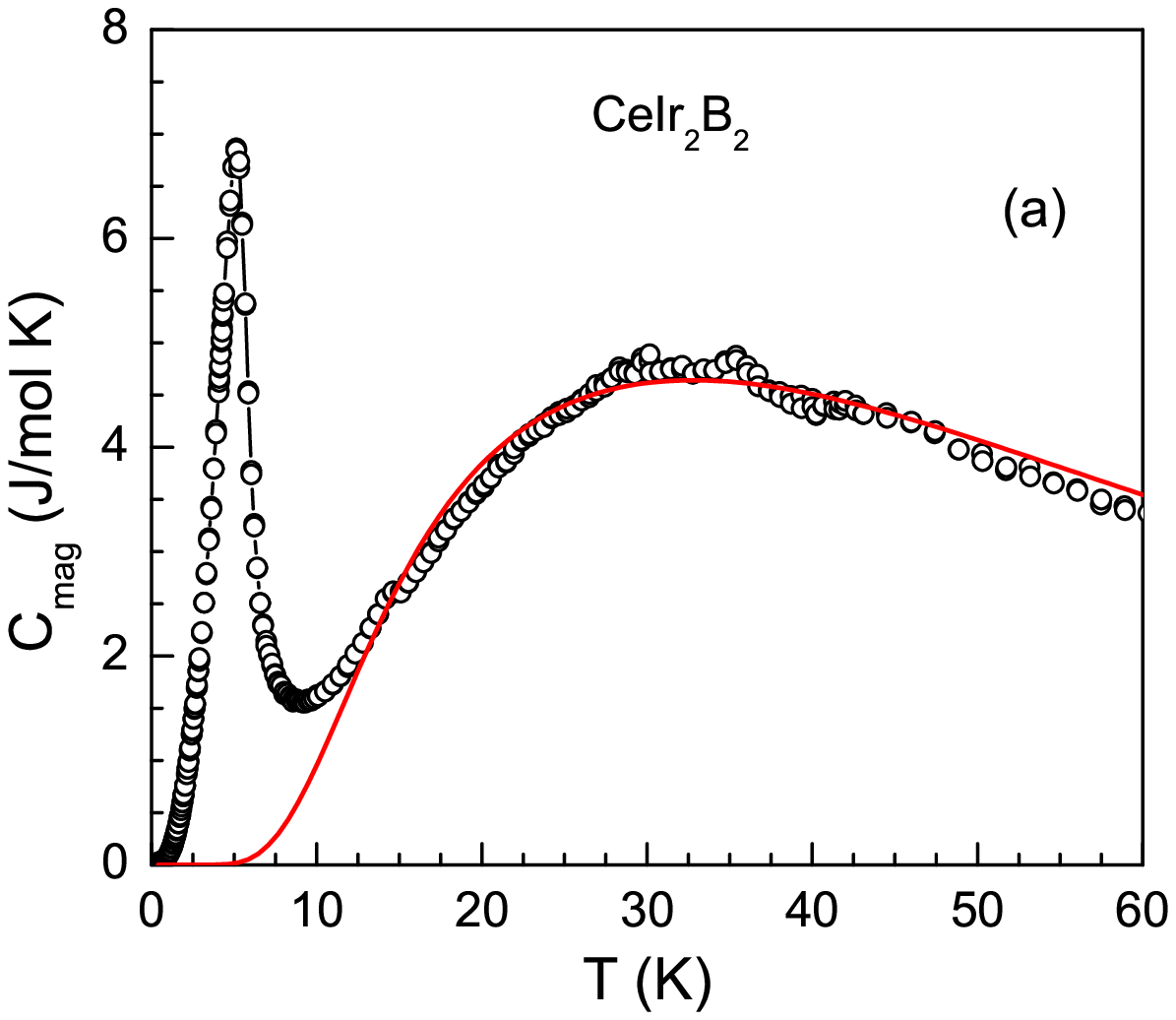}\vspace{0.1in}
\includegraphics[width=8cm, keepaspectratio]{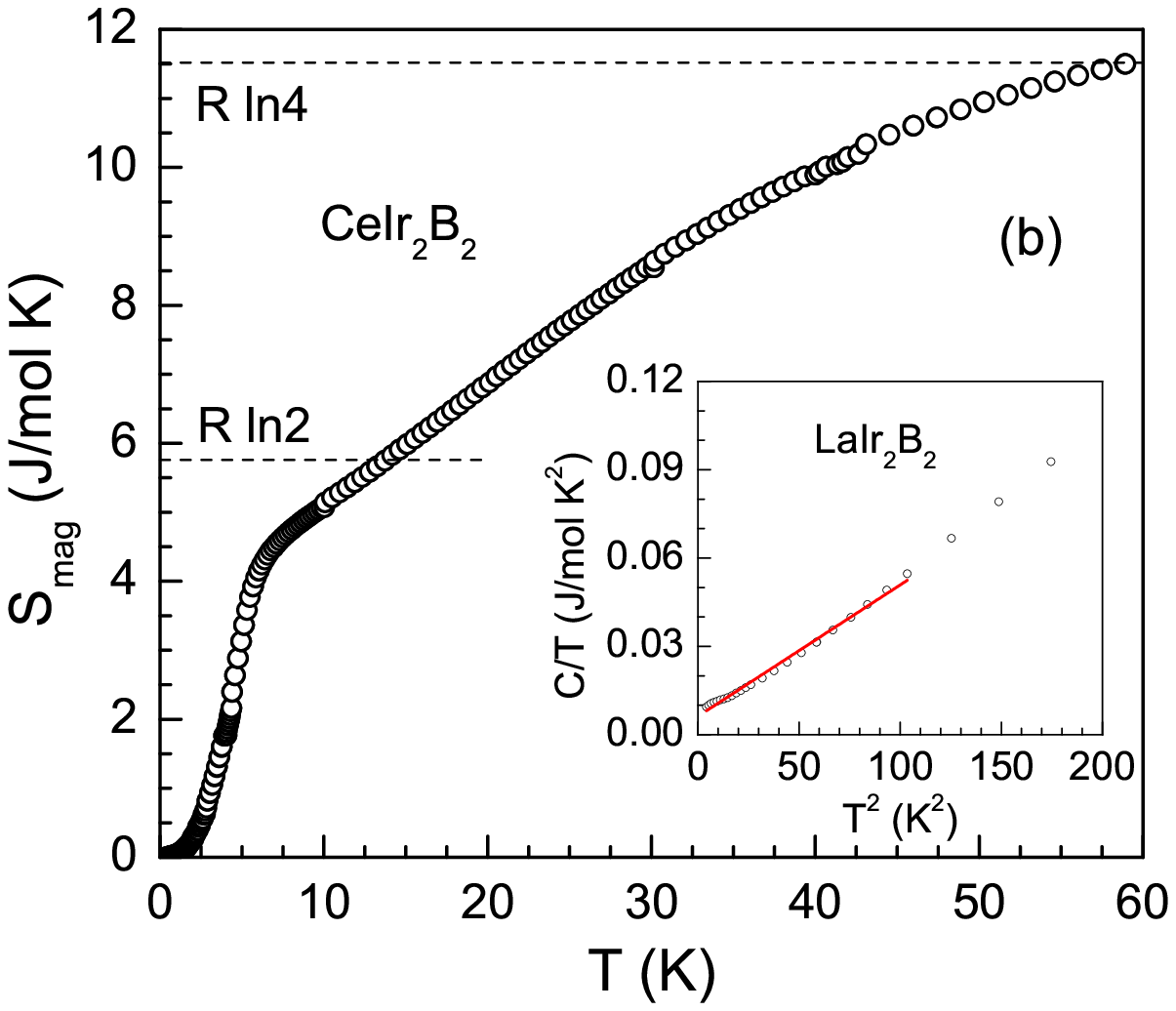}
\caption{\label{fig:CeIr2B2_Cmag} (Color online) Temperature dependence of (a) the magnetic contribution to specific heat, $C_{\rm mag}(T)$, and (b) magnetic entropy, $S_{\rm mag}(T)$ of CeIr$_{2}$B$_{2}$. The solid curve in (a) represents the fit of $C_{\rm mag}(T)$ data by Eq.~(\ref{eq:C-schottky}) for three-level CEF scheme. The inset in (b) shows the $C/T$ vs. $T^2$ plot for LaIr$_{2}$B$_{2}$.}
\end{figure}

Figure~\ref{fig:CeIr2B2_HC} shows the temperature dependence of specific heat $C(T)$ of CeIr$_{2}$B$_{2}$ and its nonmagnetic reference LaIr$_{2}$B$_{2}$. Bulk nature of ferromagnetic ordering in CeIr$_{2}$B$_{2}$ is confirmed by prominent $\lambda$-type peak in specific heat data at $T_{c} = 5.1$~K with a peak height of $\sim 7$~J/mol\,K\@. At low temperatures below $T_{c}$, the specific heat data follow
\begin{equation}
C(T) = \gamma T + \beta T^{3}+ \delta T^{3/2}exp \left(-\frac{E_g}{T}\right),
 \label{eq:C-magnon}
\end{equation}
where $\gamma T$ is the electronic contribution to the specific heat, $\beta T^{3}$ is the lattice contribution, and the last term $\delta T^{3/2}exp(-E_g/T)$ represents the spin wave contribution for a ferromagnet with an energy gap $E_g$ in the magnon-spectrum, $E_k = E_g + D k^2$, where $D$ is the spin wave stiffness constant that depends on material and $k$ is wave vector.\cite{Gopal} In order to reduce the fitting parameters we first determine $\beta$ from the low-temperature specific heat data of ${\rm LaIr_2B_2}$. An analysis of the low-temperature $C(T)$ data of ${\rm LaIr_2B_2}$ by
\begin{equation}\label{eq:HC}
  C(T) = \gamma T +\beta T^{3}
\end{equation}
in the temperature range 2~K~$\leq T \leq $~10~K gives $\gamma = 6(1)$~mJ/mol\,K$^{2}$, $\beta = 0.47(3)$~mJ/mol\,K$^{4}$\@ (see inset of Fig.~\ref{fig:CeIr2B2_Cmag}(b)). The value of the Debye temperature $\Theta_{\rm D}$ can be estimated from $\beta$ using the relation \cite{Kittel}
\begin{equation}
\Theta_{\rm D} = \left( \frac{12 \pi^{4} n R}{5 \beta} \right)^{1/3},
 \label{eq:Debye-Temp}
\end{equation}
\noindent where $R$ is molar gas constant and $n = 5$ is the number of atoms per formula unit (f.u.)\@. Thus we obtain Debye temperature $\Theta_{\rm D}$ = 274(6)~K\@ for ${\rm LaIr_2B_2}$.

The ordered state $C(T)$ data of ${\rm CeIr_2B_2}$ were fitted by by Eq.~(\ref{eq:C-magnon}) in the temperature range $0.4~{\rm K} \leq T \leq 4.5$~K which is shown by the solid curve in the inset of Fig.~\ref{fig:CeIr2B2_HC} with $\beta$ fixed to the value obtained above for ${\rm LaIr_2B_2}$. The best fit is obtained for $\gamma = 73(4)$~mJ/mol\,K$^{2}$, $\delta=1.29(7)$~J/mol\,K$^{5/2}$ and energy gap $E_g=3.65(4)$~K\@. The Sommerfeld coefficient $\gamma = 73(4)$~mJ/mol\,K$^{2}$ obtained this way is found to be less than the reported value of 180~mJ/mol\,K$^{2}$ obtained from the extrapolation of the $C(T)$ data in the paramagnetic state in Ref.~\onlinecite{Sampatkumaran}. However, in comparison with other ferromagnetic systems like CeRu$_{2}$Ge$_{2}$ ($\gamma=20$~mJ/mol\,K$^{2}$) \cite{Bohm, Wilhelm} and CePd$_{2}$Ga$_{3}$ ($\gamma=9$~mJ/mol\,K$^{2}$) \cite{Bauer} this $\gamma$ value is rather large. Usually in Ce-based heavy fermion systems $\gamma$ value are found to be larger than 100~mJ/mol\,K$^{2}$. \cite{Kadowaki, Tsujii} Thus the $\gamma = 73(4)$~mJ/mol\,K$^{2}$ characterizes ${\rm CeIr_2B_2}$ as a moderate heavy fermion system.

The density of states at the Fermi level ${\cal D}(E_{\rm F})$ can be estimated from Sommerfeld coefficient $\gamma$ using the relation \cite{Kittel}
\begin{equation}
\gamma = \frac{\pi^2 k_{\rm B}^2}{3} {\cal D}(E_{\rm F}),
\label{eq:DOS}
\end{equation}
\noindent where $k_{\rm B}$ is Boltzmann's constant. Using $\gamma = 6(1)$~mJ/mol\,K$^{2}$ we obtain ${\cal D}(E_{\rm F}) = 2.54~{\rm states/(eV~f.u.)}$ for both spin directions for LaIr$_{2}$B$_{2}$. The renormalization factor for quasi-particle density of states due to $4f$ correlations in CeIr$_{2}$B$_{2}$ is
\begin{equation}\label{Eq:gamma-renorm}
\frac{\gamma({\rm CeIr_2B_2})}{\gamma({\rm LaIr_2B_2})} = \frac{73}{6} \approx 12.
\end{equation}
Thus the renormalized quasi-particle mass in  CeIr$_{2}$B$_{2}$ is $m^* \approx  12 \,m_{\rm e}$, where $m_{\rm e}$ is the free electron mass.

Figure~\ref{fig:CeIr2B2_Cmag} shows the magnetic contributions to the specific heat $C_{\rm mag}(T)$ and entropy $S_{\rm mag}(T)$ of CeIr$_{2}$B$_{2}$. The magnetic contribution to the specific heat of CeIr$_{2}$B$_{2}$ was obtained by subtracting the specific heat of LaIr$_{2}$B$_{2}$ assuming the lattice contribution to be approximately equal to that of LaIr$_{2}$B$_{2}$, and the magnetic entropy $S_{\rm mag}(T)$ was obtained by integrating the $C_{\rm mag}(T)/T$ vs. $T$ plot. Apart from the sharp $\lambda$-type peak of ferromagnetic transition we also observe a broad Schottky-type anomaly centered around 30~K in $C_{\rm mag}(T)$. The Schottky-type anomaly could be reproduced by the analysis of $C_{\rm mag}(T)$ data by three-level crystal electric field scheme. For a three CEF level system, Schottky contribution to the specific heat is given by \cite{Layek}
\begin{eqnarray}
C_{\rm Sch}(T) &=& R\,\Bigg[g_0g_1 \left(\frac{\Delta_1}{T}\right)^2 e^{-\frac{\Delta_1}{T}} + g_0g_2 \left(\frac{\Delta_2}{T}\right)^2 e^{-\frac{\Delta_2}{T}} \nonumber \\
     & & \hspace{1cm} +~g_1g_2\left(\frac{\Delta_1-\Delta_2}{T}\right)^2 e^{-\frac{\Delta_1 + \Delta_2}{T}} \Bigg] \nonumber\\
     & & \hspace{1cm} \times~ \Bigg[ \frac{1}{g_0 + g_1 e^{-\frac{\Delta_1}{T}}+ g_2 e^{-\frac{\Delta_2}{T}}} \Bigg]^2
\label{eq:C-schottky}
\end{eqnarray}
where $g_0$, $g_1$ and $g_2$ are the degenracies of the ground state, first excited state and second excited state, respectively, and $\Delta_1$ and $\Delta_2$ are the splitting energies between the ground state and the first excited state, and between the ground state and the second excited state, respectively. In an orthorhombic environment the $(2J+1)$-fold degenerate ground state multiplet of Ce$^{3+}$ ion $(J=5/2)$ splits into three doublets, thus $g_0 = g_1 = g_2 = 2$\@. The $C_{mag}(T)$ data in Fig.~\ref{fig:CeIr2B2_Cmag}(a) were fitted by Eq.~(\ref{eq:C-schottky}). The least squares fit of $C_{mag}(T)$ data by Eq.~(\ref{eq:C-schottky}) is shown by the solid curve in Fig.~\ref{fig:CeIr2B2_Cmag}(a) which was obtained for $\Delta_1 = 56.1(3)$~K and $\Delta_2 = 140.8(9)$~K\@. The temperature dependence of $S_{\rm mag}(T)$ in Fig.~\ref{fig:CeIr2B2_Cmag}(b) indicates that the magnetic entropy attains a value $S_{\rm mag} = R\ln4 $ at 60~K, which further supports the splitting energy of 56~K between the ground state and the first excited state determined above.

Further, we observe that even at 6.5~K, which is above the $T_{c}$, the magnetic entropy $S_{\rm mag} \sim 0.75 R\ln 2$ which is considerably lower than that of $R\ln2$ expected for a magnetic doublet ground state. The reduced value of $S_{mag}$ together with moderately large value of $\gamma$ suggests that the Kondo effect is weak but sizable in this compound and the RKKY interaction dominates leading to a magnetically ordered state below 5.1~K. The Kondo temperature $T_{\rm K}$ can be estimated from the mean field theoretical universal plot of jump in specific heat $\Delta C_{\rm mag}$ vs. $T_{\rm K}/T_m$ obtained by Besnus et al. \cite{Besnus} and Blanco et al. \cite{Blanco} for Ce-based Kondo lattice systems. For the observed $\Delta C_{\rm mag} = 6.16$~J/mol\,K at $T_c = 5.1$~K for our compound we find $T_{\rm K}/T_{c} \approx 0.83$ from this universal plot. Thus we obtain $T_{\rm K} \approx  4.2$~K\@. The Kondo temperature can also be estimated from the Weiss temperature, $T_{\rm K} \approx |\theta_{\rm p}|/4.5 $, \cite{Gruner} which using $\theta_{\rm p} = -13.4(2)$~K yields $T_{\rm K} \approx  3$~K\@ which is consistent with the above estimate of $T_{\rm K}$ from the specific heat jump at $T_c$\@.

\begin{figure}
\includegraphics[width=8cm, keepaspectratio]{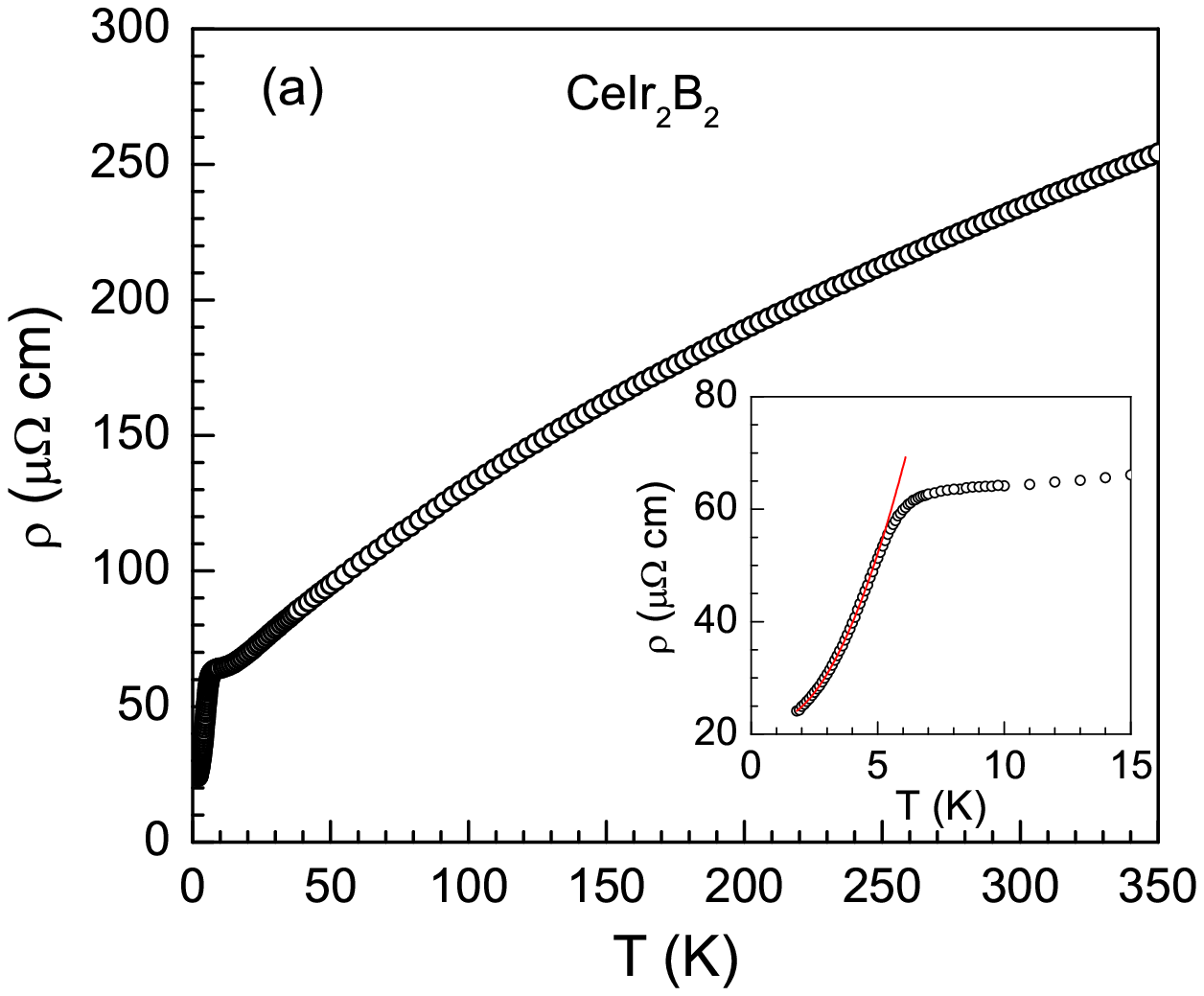}\vspace{0.1in}
\includegraphics[width=8cm, keepaspectratio]{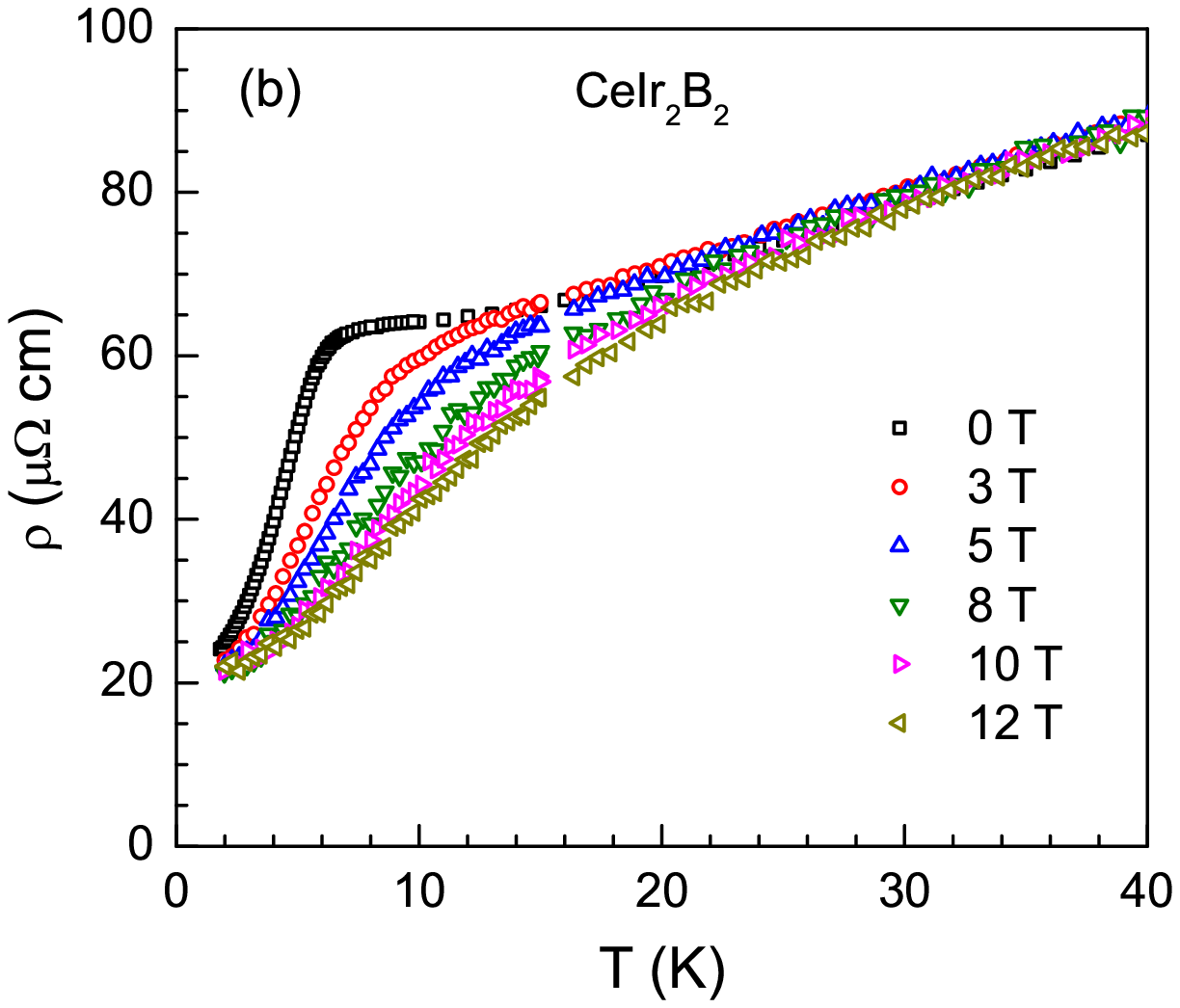}
\caption{\label{fig:CeIr2B2_Rho} (Color online) (a) Temperature dependence of electrical resistivity data, $\rho(T)$ of CeIr$_{2}$B$_{2}$ measured in zero field in the temperature range (2--300~K). The inset shows an expanded view of the low temperature $\rho(T)$ data. The solid curve in inset represents a fit to gapped magnon behavior by Eq.~(\ref{eq:rho-magnon}) for $2.0~{\rm K} \leq T \leq 4.5$~K. (b) The low temperature $\rho(T)$ data measured under different applied field $H$.}
\end{figure}

The electrical resistivity $\rho(T)$ of CeIr$_{2}$B$_{2}$ as shown in Fig.~\ref{fig:CeIr2B2_Rho}(a) exhibits a metallic behavior with residual resistivity, $\rho_0 =  24.1 ~\mu \Omega$\,cm at 2~K and residual resistivity ratio, RRR = $\rho_{300\,{\rm K}}/\rho_{2\,{\rm K}} \sim  10$. The low value of $\rho_0$ and high RRR indicate the good quality of our sample. We observe a very broad curvature around 150~K in $\rho(T)$ which may be the result of the combined effect of crystal electric field and Kondo interaction. According to the single impurity Kondo model proposed by Cornut and Coqblin, \cite{Cornut} the position of the high temperature maximum in $\rho(T)$ gives a rough estimate of the total CEF splitting. Thus from resistivity the overall CEF splitting enegy should be around $\sim 150$~K which is consistent with the above obtained overall splitting of $\sim 141$~K from specific heat data. The overall behavior of $\rho(T)$ is in agreement with that reported by Sampathkumaran et al. \cite{Sampatkumaran}  As can be seen from the inset of Fig.~\ref{fig:CeIr2B2_Rho}(a), there is a sharp  transition at 6~K in $\rho(T)$ due to the onset of magnetic order, below 6~K $\rho$ drops very rapidly due to the reduction in spin disorder scattering. In the ordered state, similar to specific heat data, the $\rho(T)$ data also reveal a gapped-magnon behavior.

For a ferromagnet, the spin-wave contribution to resistivity due to electron-magnon scattering leads to $\rho_{\rm m} \sim T^2$ temperature dependence. However, if there is an energy gap in the magnon spectrum then electron-magnon resitivity $\rho_{\rm m} $ is given by \cite{Anderson1979}
\begin{equation}
\rho_{\rm m} = BT\frac{E_g}{k_{\rm B}} \left(1+\frac{2k_{\rm B}T}{E_g}+\frac{1}{2}e^{-E_g/k_{\rm B}T}+...\right)e^{-E_g/k_{\rm B}T},
\label{eq:rho-magnon}
\end{equation}
where
\begin{equation}
B = \frac{1}{4(J+1)}\left(\frac{k_{\rm B}D}{k_{\rm F}^2}\right)^2\rho_{\rm sd}.
\label{eq:constantB}
\end{equation}
 The spin-disorder resistivity $\rho_{\rm sd}$ due to the presence of disordered magnetic moments is given by \cite{Anderson1974}
\begin{equation}
\rho_{\rm sd} = \frac{3\pi}{8}N_{\rm ion}\frac{A^2(g-1)^2m_{\rm e}}{e^2\hbar E_{\rm F}} J(J+1),
\label{eq:rhosd}
\end{equation}
where $N_{\rm ion}$  is the number of Ce$^{3+}$ ions per unit volume, $A$ is the interaction strength, $g$ is the gyromagnetic factor, $e$ is the fundamental electronic charge, $m_{\rm e}$ is the the free electron mass, and $E_{\rm F}$ is the Fermi energy.

The $\rho(T)$ data below 4.5 K were fitted by
\begin{equation}
\rho(T) = \rho_{0} + BT\frac{E_g}{k_{\rm B}} \left(1+\frac{2k_{\rm B}T}{E_g}+\frac{1}{2}e^{-E_g/k_{\rm B}T}\right)e^{-E_g/k_{\rm B}T},
\label{eq:rho-magnon-fit}
\end{equation}
where $\rho_0$ is the residual resistivity due to the scattering by static defects in crystal lattice. The solid curve in the inset of Fig.~\ref{fig:CeIr2B2_Rho}(a) shows the fit of $\rho(T)$ data by Eq.~(\ref{eq:rho-magnon-fit})\@. The best fit was obtained for $\rho_{0} = 22.1(1)~\mu\Omega$\,cm, $B =0.76(1)~\mu\Omega$\,cm/K$^{2}$ and $E_g = 2.7(1)$~K\@. This value of energy gap $E_g =2.7(1)$~K is slightly smaller but comparable to the value obtained from the analysis of ordered state $C(T)$ data above. The existence of an energy gap in the magnon spectrum as evidenced by low temperature $\rho(T)$ and $C(T)$ data indicate that the magnetic properties are highly anisotropic in ordered state. Ferromagnetic system CeNiIn$_{2}$, which has $T_{c}=3.4$~K, also shows similar gapped-magnon behavior in specific heat and resistivity in the magnetically ordered state. \cite{Rojas}

\begin{figure}
\includegraphics[width=8cm, keepaspectratio]{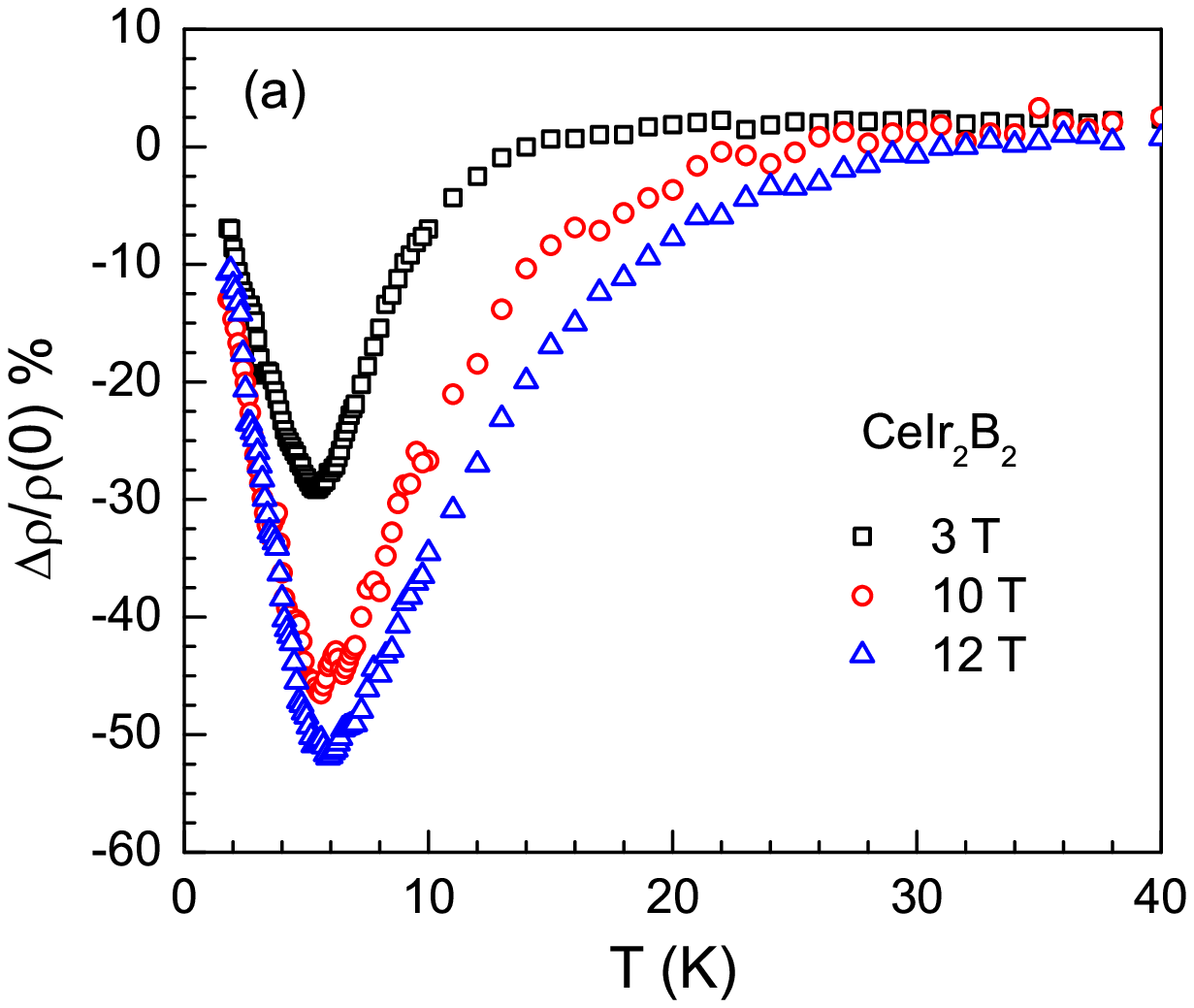}\vspace{0.1in}
\includegraphics[width=8cm, keepaspectratio]{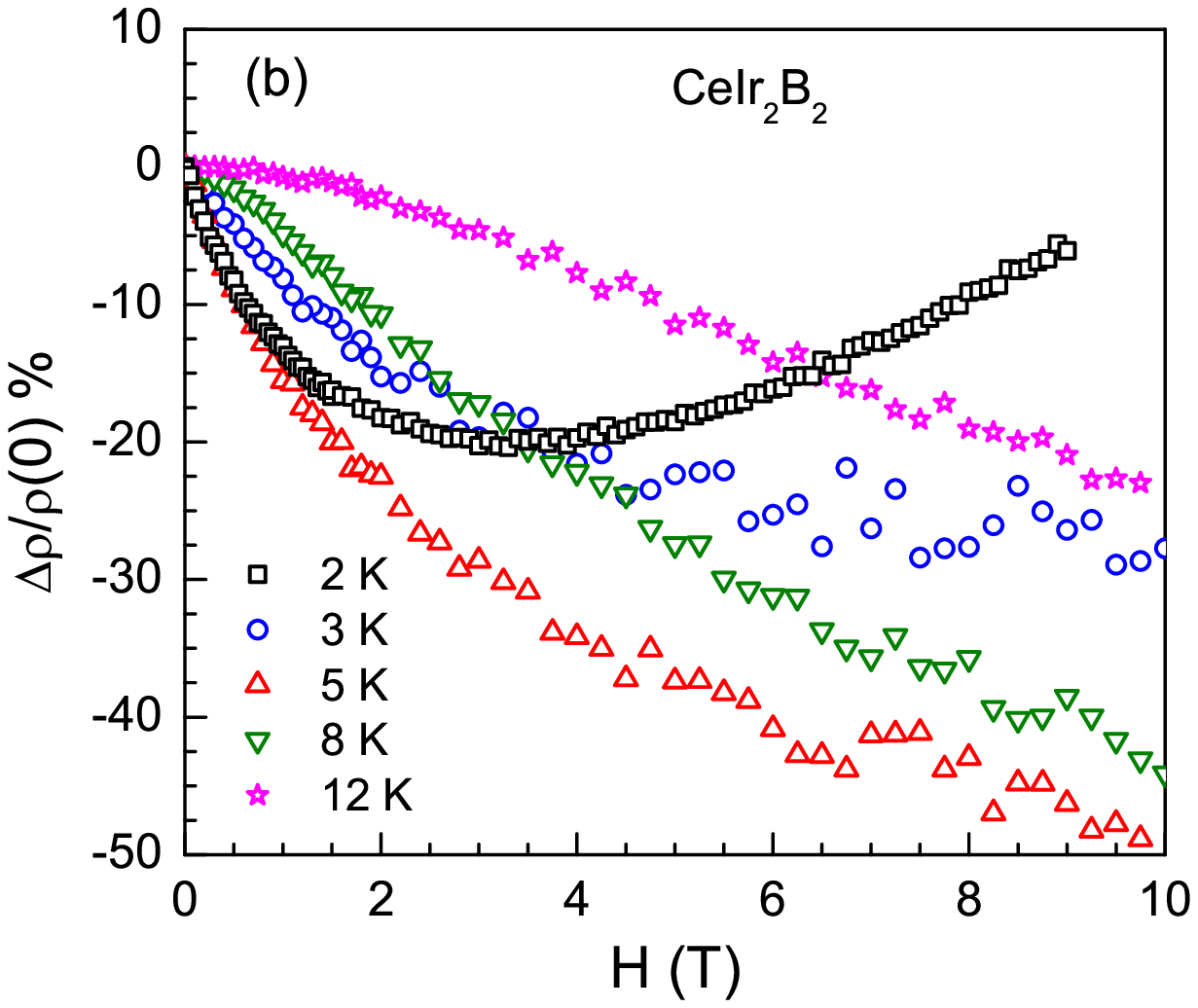}
\caption{\label{fig:CeIr2B2_MR} (Color online) The magnetoresistance (MR) $[\Delta\rho(H)/\rho(0)]$~\% for CeIr$_{2}$B$_{2}$ (a) as function of temperature $T$ at indicated magnetic fields, and (b) as a function of magnetic field $H$ at indicated temperatures. Note: The 2~K MR data in (b) were collected on a different sample than all other MR data presented in this figure.}
\end{figure}

The $\rho(T)$ data measured under various applied fields are shown in Fig.~\ref{fig:CeIr2B2_Rho}(b). It is seen that the increase in applied field smoothens and broadens the $\rho(T)$ anomaly near $T_{c}$. With an increase in applied field the transition temperature is also seen to shift towards the higher temperature side which is a characteristic of ferromagnetic ordering. Further, we observe that the application of magnetic field causes a very small decrease in residual resistivity leading to a negative magnetoresistance as expected for a ferromagnetically ordered system. The magnetoresistance (MR), $\Delta\rho(H)/\rho(0)$ calculated from the $\rho(T)$ data measured at $H = 3$~T, 8~T and 12~T are shown in Fig.~\ref{fig:CeIr2B2_MR}(a). The MR is defined as
\begin{equation}
\frac{\Delta \rho(H)}{\rho(0)} = \frac{\rho(H)- \rho(0)}{\rho(0)}
 \label{eq:MR}
\end{equation}
where $\rho(H)$ is the resistivity measured at an applied field $H$\@. It is seen from the $T$ dependence of MR that at 3~T the MR is negative and decreases with increasing $T$, shows a minima near $T_{c}$, and starts increasing above that and eventually becomes positive above 14~K\@. The overall trend of MR remains same even at higher fields e.g., at 8~T and 12~T, however the temperature range above $T_{c}$ over which the MR is negative is found to be larger at the higher fields. The MR has highest absolute value near $T_{c}$, which is about 52 \% at 12~T.

We have also investigated the magnetic field dependence of the electrical resistivity, $\rho(H)$, which is displayed in Fig.~\ref{fig:CeIr2B2_MR}(b) as magnetoresistance. In the magnetically ordered state, at 2 K, it is seen from Fig.~\ref{fig:CeIr2B2_MR}(b) that the MR is initially negative as expected for a ferromagnetic system. As the magnetic field is increased the MR first decreases rapidly, exhibits a minimum around $\sim 3.5$~T (MR $\sim -20 $ \%) and then again rises continuously with applied field up to the investigated field of 9~T\@. The upturn in MR above $\sim 3.5$~T at 2~K can be understood to result from two different contributions: (i) negative MR arising from the reduction in spin disorder resistivity that gets rapidly suppressed by magnetic field due to polarization, and (ii) positive MR due to classical modification of trajectory by Lorentzian force. The second contribution which is proportional to $H^2$ wins over the first contribution at higher field above $\sim 3.5$~T, resulting in the observed upturn in MR. However, as the temperature is increased, i.e. at 3~K, the positive MR contribution is reduced leading to a negative MR throughout up to 10~T. At 5 K which is very close to $T_c$, the absolute value of MR is maximum (a negative MR of $\sim -50$ \% at 5~K and 10~T)\@. This behavior of MR at 5~K is similar to that observed by Sampathkumaran et al. \cite{Sampatkumaran} however with a slightly smaller value than the reported one. Since $T= 5$~K is very close to the boundary separating the paramagnetic and ferromagnetic phases, therefore MR might have negative contributions from both the phases coexisting together at 5~K\@ yielding the maximum absolute MR at this temperature. At the further higher temperatures the MR though negative throughout has smaller values. Even at 12~K we observe a negative MR of $\sim -23$ \% at 10~T\@. Such a high MR in paramagnetic state may suggest the presence of ferromagnetic correlations well above the ordering temperature consistent with the NMR results discussed below.

\begin{figure}
\includegraphics[width=8cm, keepaspectratio]{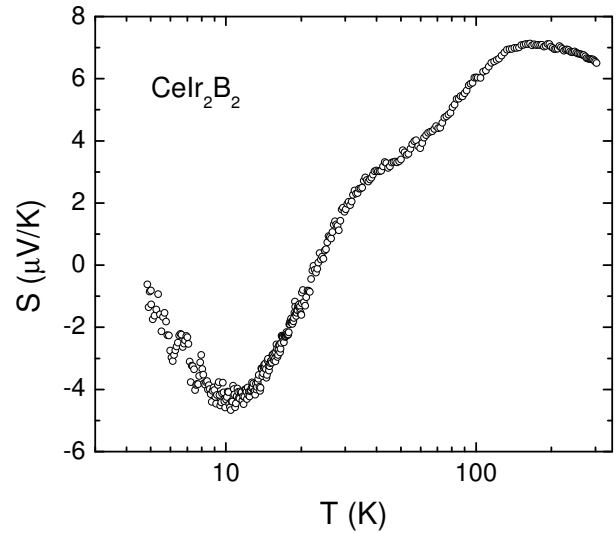}
\caption{\label{fig:CeIr2B2_Thermopower} (Color online) Temperature dependence of thermoelectric power, $S(T)$ of CeIr$_{2}$B$_{2}$ measured in zero field.}
\end{figure}

The data from the thermoelectric power $S(T)$ measurements of CeIr$_{2}$B$_{2}$ sample are shown in Fig.~\ref{fig:CeIr2B2_Thermopower}. The $S(T)$ exhibits a shoulder around 30 K and a maximum around 150 K with an absolute value of 7 $\mu$V/K, which is small compared to typical Ce-based Kondo systems. The Kondo systems usually exhibit a large absolute values of $S\sim$~10--100~$\mu$V/K.\cite{Garde, Zatic} At  low temperatures $S(T)$ changes its sign  at $\sim 25$~K and shows a minimum at 10 K with $S = - 5~\mu$V/K and again starts to increase with decreasing temperatures. The minimum just above the transition temperature is also seen for many magnetically ordered compounds for example in ferromagnetic CeGe$_{2}$ ($T_{c} = 7$~K), for which a negative peak is observed at around 10 K just above $T_{c}$.\cite{Jaccard1990} The high-$T$ anomalies are likely connected with Kondo scattering on the two excited CEF doublets. The low absolute value of the thermoelectric power, however, suggests that the Kondo interactions are weak in CeIr$_{2}$B$_{2}$.

\begin{figure}
\includegraphics[width=8cm, keepaspectratio]{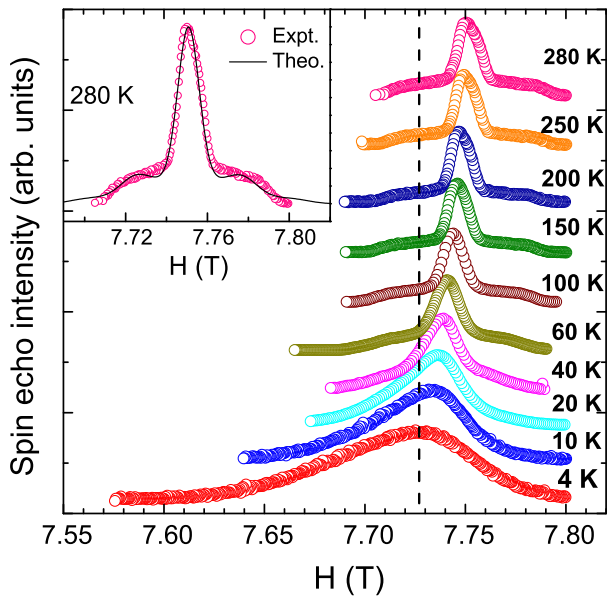}
\caption{\label{fig:CeIr2B2_NMR} (Color online) $^{11}$B field sweep NMR spectra at various constant temperatures. The vertical line indicates the reference position with $^{11}K=0$ obtained from the reference compound H$_{3}$BO$_{3}$. The inset shows a typical powder pattern at 280~K together with the simulation.}
\end{figure}

Figure~\ref{fig:CeIr2B2_NMR} shows the $^{11}$B $(I=3/2)$ field sweep NMR spectra at different temperatures. At high temperature the spectra are typical powder pattern with moderate broadening and first order quadrupolar splitting. Despite the orthorhombic structure the spectra are fitted well assuming one B-lattice site with isotropic shift and broadening. A quadrupolar coupling constant of $\nu_{Q}= 0.89$~MHz is obtained from the fitting. The vertical dotted line indicates the position of the Larmor field obtained from the reference compound H$_{3}$BO$_{3}$ under similar NMR conditions. With decreasing temperature the spectra are shifted towards the low field side, i.e., approaching towards the nonmagnetic reference line, with gradual line broadening. Furthermore a sizable anisotropy shows up in the spectra. The total Knight shift has two components $^{11}K$ = $^{11}K_{0}$ + $^{11}K_{4f}$. $^{11}K_{0}$ is negative and temperature independent, whereas $^{11}K_{4f}$ is positive and temperature dependent. $^{11}K_{4f}$ denotes the contribution due to conduction electron polarization by $4f^{1}$ Ce-ions. We shall discuss $^{11}K_{4f}$ in the following paragraph. The broadening of the spectra is rather pronounced below 20~K. Moreover the spectra appear to be more anisotropic below this temperature.  This broadening is related with the onset of long range ferromagnetic ordering at $T_{c} = 5.1$~K. The additional broadening below 20~K indicates that the $^{11}$B nuclei are able to sense the ferromagnetic correlation well above the ordering temperature.

\begin{figure}
\includegraphics[width=8cm, keepaspectratio]{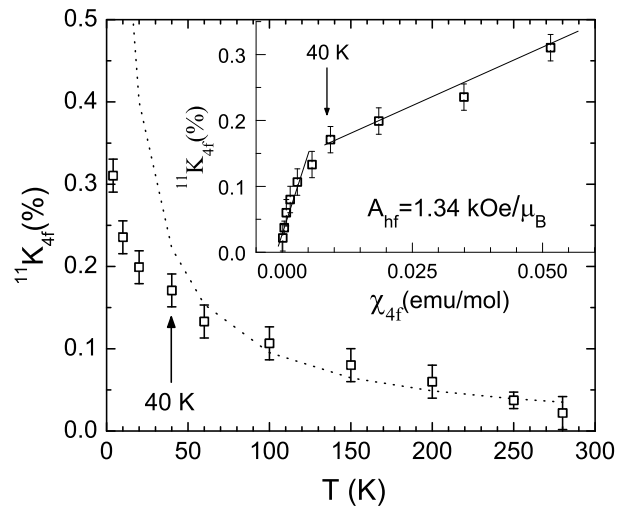}
\caption{\label{fig:CeIr2B2_NMR-K} (Color online)  $^{11}K_{4f}$  as a function of temperature. The dotted line indicates a Curie-Weiss law.  The inset shows the $^{11}K_{4f}$ vs. $\chi_{4f}$  plot.}
\end{figure}

Figure~\ref{fig:CeIr2B2_NMR-K} shows the temperature dependence of the $4f$ contribution to the Knight shift, $^{11}K_{4f}$. Although $^{11}K_{4f}(T)$ continuously increases with decreasing temperature, it does not follow a simple Curie-Weiss law (visualized by the dotted line).  A closer inspection by plotting $^{11}K_{4f}(T)$ versus the bulk susceptibility reveals two different regimes, one above and one below 40~K, respectively. In both regimes the dependence is close to linear, but the slopes differ by at least a factor of five. The calculated hyperfine coupling constant $A_{\rm hf}$ from the $^{11}K_{4f}$ vs. $\chi_{4f}$ plot in the high temperature range is around 1.35~kOe/$\mu_{\rm B}$. This value is smaller than the value of $^{29}$Si $A_{\rm hf}$ of heavy fermion systems CeRu$_2$Si$_2$ and CeCu$_2$Si$_2$, respectively.\cite{Aarts1983, Matsuda2000} The low-$T$ regime corresponds to the region where the huge broadening of the spectra takes place. Therefore the deviation from the Curie-Weiss behavior and the break in the slope of  $^{11}K_{4f}(T)$ versus $\chi_{4f} (T)$ plot are very likely related to the onset of ferromagnetic correlations. There one expects ordering along a specific direction which results in a large anisotropy of the bulk magnetization and a large anisotropy of the internal field. In such a case the peak position in the NMR spectra is not directly related to the powder average of the bulk susceptibility. The non-linearity at high temperatures is likely related to CEF effects, which are expected to induce a temperature dependence in the anisotropy of the susceptibility and departure from a strict Curie-Weiss behavior.

\section{\label{Conclusions} CONCLUSIONS}

To summarize, our comprehensive studies by $\chi(T)$, $C(T)$, $\rho(T, H)$, $S(T)$, and $^{11}$B NMR measurements confirm the presence of long range magnetic ordering in CeIr$_{2}$B$_{2}$. Ce ions in CeIr$_{2}$B$_{2}$ are in stable trivalent states which undergo ferromagnetic ordering at $T_c =5.1$~K\@. Further, the observation of a gapped-magnon behavior in both $C(T)$ and $\rho(T)$ data below $T_c$ indicates that the magnetic properties of this compound are highly anisotropic. A CEF analysis of $C_{\rm mag}(T)$ data suggests that the ground state is a CEF doublet with an overall crystal field splitting energy of $\sim 141$~K\@. The Kondo interaction is rather weak as evidenced from the low absolute values of the thermoelectric power and  moderately enhanced value of the Sommerfeld coefficient $\gamma = 73(4)$~mJ/mol\,K$^{2}$\@. As a result of the dominant RKKY interactions the system orders magnetically. A shoulder at 30~K and a maximum at around 150 K in $S(T)$ are likely connected with the excited CEF doublets. NMR results indicate that the $^{11}$B nuclei senses the ferromagnetic correlation well above the ordering temperature. Despite the orthorhombic structure, the NMR spectra above 40~K can be simulated without significant anisotropy. Below 40~K the NMR spectra broadens considerably, which is attributed to the onset of ferromagnetic correlations. This is likely also the origin of a strong deviation of the $T$-dependence of the $^{11}K_{4f}$ shift from a Curie-Weiss behavior and of a pronounced change in the $^{11}K(T)$ versus $\chi(T)$ dependence. The presence of ferromagnetic correlations well above the ordering temperature is also observed in MR data. The $T$ dependence of MR shows a minima near $T_c$ with a large negative MR of $\sim -52$~\% at 12~T. Further investigations such as neutron diffraction and pressure studies are required to determine the magnetic structure and suppress the magnetic order and investigate the pressure induced quantum phase transition.

\section*{ACKNOWLEDGEMENTS}
Financial assistance from BRNS, Mumbai and Indian Institute of Technology, Kanpur is acknowledged. VKA acknowledges support from the US Department of Energy-Basic Energy Sciences under Contract No.~DE-AC02-07CH11358

\end{document}